\documentclass[9pt,journal]{IEEEtran}

\IEEEoverridecommandlockouts
\usepackage{color}

\usepackage{amsthm,amsmath,amssymb}
\usepackage{mathtools}

\usepackage{verbatim}
\usepackage{cite}
\usepackage{url}
\usepackage{cases}
\usepackage{stmaryrd}

\let\labelindent\relax
\usepackage[inline]{enumitem}

\usepackage[flushleft]{threeparttable} 

\usepackage[named]{algo}
\usepackage[noend]{algpseudocode}
\usepackage{algorithm}

\usepackage{widetext}

\usepackage{upgreek}
\usepackage{bm}
\usepackage{nicefrac}

\ifCLASSOPTIONcompsoc
  \usepackage[caption=false,font=normalsize,labelfont=sf,textfont=sf]{subfig}
\else
  \usepackage[caption=false,font=footnotesize]{subfig}
\fi

\usepackage{cite}
\usepackage{balance}

\usepackage{lipsum}

\newtheorem{theorem}{Theorem}

\newtheorem{lemma}[theorem]{Lemma}

\IEEEoverridecommandlockouts

\begin{document}

\title{RaSSteR: Random Sparse Step-Frequency Radar}

\author{Kumar Vijay~Mishra, Satish~Mulleti and Yonina C.~Eldar
\thanks{K. V. M. is with The University of Iowa, Iowa City, IA 52246 USA, e-mail: kumarvijay-mishra@uiowa.edu.}
\thanks{S. M. and Y. C. E. are with the Faculty of Mathematics and Computer Science, Weizmann Institute of Science, Rehovot 7610001 Israel, e-mail: \{satish.mulleti, yonina.eldar\}@weizmann.ac.il.}
\thanks{This project has received funding from the European Union’s Horizon 2020 research and innovation program under grant No. 646804-ERC-COG-BNYQ, Air Force Office of Scientific Research under grants No. FA9550-18-1-0208, and from the Israel Science Foundation under grant No. 0100101.}
}

\maketitle

\begin{abstract}
We propose a method for synthesizing high range resolution profiles (HRRP) using stepped frequency waveform (SFW) processing. Conventional SFW radars sweep over the available spectrum linearly to achieve high resolution from their instantaneous bandwidth. However, they suffer from strong range-Doppler coupling and coexisting spectral interference. Prior works are able to mitigate only one of these drawbacks. We present a new \textit{ra}ndom \textit{s}parse \textit{ste}p-frequency \textit{r}adar (RaSSteR) waveform that consumes less spectral resources without loss of range resolution and estimates both high-resolution range and Doppler by exploiting sparse recovery techniques. In the presence of interference, the operation with the new waveform is made cognitive by focusing available transmit power only in the few transmit bands. Our theoretical analyses show that, even while using fewer carriers in the available bandwidth, RaSSteR has identical recovery guarantees as the standard random stepped frequency (RSF) waveform. Numerical experiments demonstrate performance enhancements with RaSSteR over state-of-the-art such as SFW, RSF, conventional pulse-compression-based pulse Doppler radar, and sub-Nyquist radar. In addition, the target hit rate of RaSSteR in the presence of strong interference is $30$\% more than conventional RSF.
\end{abstract}


\IEEEpeerreviewmaketitle

\section{Introduction}
\label{sec:intro}
In many remote sensing problems, obtaining information at high range resolution, namely the ability to distinguish closely located targets, is of paramount importance \cite{levanon2004radar}. Typical applications include cases in which the objects are distant as in synthetic aperture radar (SAR) and inverse SAR (ISAR) \cite{soumekh1999synthetic}; hidden as in through-wall imaging and foliage penetration (FOPEN) radar \cite{davis2011foliage}; complex as in ground penetration radar (GPR) \cite{giovanneschi2019dictionary}; and seemingly similar as in gesture radar \cite{lien2016soli}. High resolution measurements significantly aid in modeling, analysis, detection, and classification of radar targets, but also lead to increased radar system complexity and design requirements \cite{peebles1998radar}. In \textit{pulse Doppler radar} (PDR) systems, the resolution is inversely proportional to the support of the \textit{ambiguity function} of the transmit pulse. As a result, increase in the transmit signal bandwidth leads to higher range resolution \cite{mishra2019sub}. 

There is a large body of literature on classical techniques for radar transmit waveform design in order to achieve high resolution sensing \cite{gini2012waveform,rihaczek1996principles}. Broadly, they are grouped into three classes. In the first, \textit{ultra-wideband} (UWB) radar \cite{taylor2000ultra} is employed to transmit extremely narrow unmodulated pulses on the order of several nanoseconds to tens of picoseconds, thus directly resulting in a large \textit{instantaneous} bandwidth $B$; then, the range resolution $\Delta r = c/2B$, where $c=3\times 10^8$ m/s is the speed of the light. However, this imposes stringent constraints on the analog-to-digital converter (ADC) employed \cite{eldar2015sampling}. 

UWB systems require wideband radio-frequency (RF) devices that are more expensive than those used in traditional narrowband radars \cite{taylor1995introduction}. An alternative is \textit{intrapulse} pulse compression (PC), modulation inside the transmit pulse through frequency or phase coding which results in a large bandwidth \cite{lewis1986aspects}. The received pulses are processed using a digital matched filter \cite{george2010implementation}, which correlates the sampled received signal with a replica of the transmit signal resulting in a compressed pulse in the range-Doppler plane. The instantaneous bandwidth of the pulse compression signal is also large, thereby requiring high-rate ADCs. In recent years, sub-Nyquist radars \cite{mishra2019sub} have demonstrated signal detection and parameter estimation from much fewer measurements than Nyquist intrapulse PC systems.

\begin{figure*}
\centering
\includegraphics[width=0.9\textwidth]{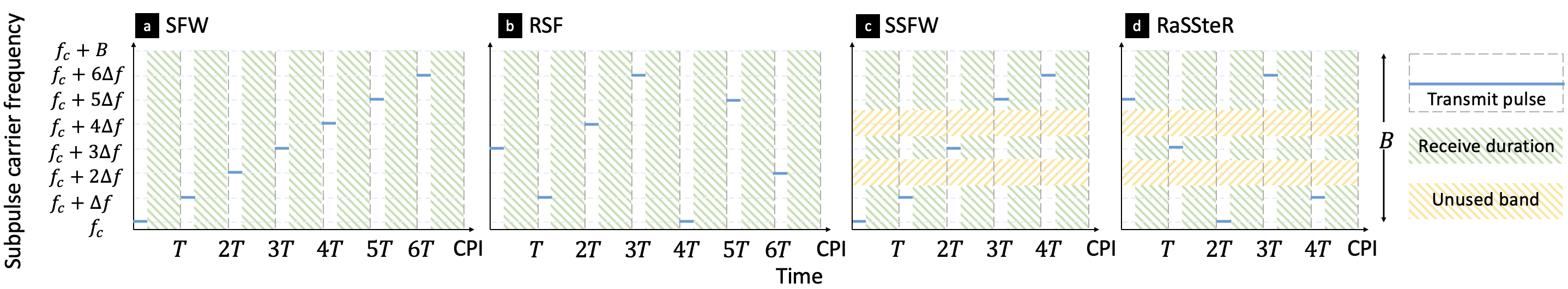}
\caption{Simplified illustration of variants of SF transmission 
for a total of $M=7$ possible carrier frequencies. (a) SFW 
(b) RSF 
(c) SSFW 
(d) RaSSteR. 
}
\label{fig:waveforms}
\end{figure*}

The third approach employs step-frequency waveform (SFW), which can be viewed as a form of \textit{interpulse} PC \cite{gill1995step}. Here, the transmit signal bandwidth is increased sequentially (Fig.~\ref{fig:waveforms}a), unlike instantly as in UWB and intrapulse PC, over several $N$ pulses. The cumulative duration of all pulses is a \textit{burst} or \textit{coherent processing interval} (CPI) $NT$, where $T$ is the pulse repetition interval (PRI). The carrier frequency of successive pulses is increased \textit{linearly} over a fixed spectral range comprising $M=N$ unique frequencies spread over the bandwidth $B$ \cite{levanon2004radar}. At the receiver, a coherent detector measures both the in-phase (I) and quadrature (Q) components of the received signal corresponding to each frequency step. The phase information embedded in this complex received signal sequence yields accurate range measurements. Hence, by processing all pulses together, the SFW radar indirectly achieves wide bandwidth and keeps the single-pulse instantaneous bandwidth low. The effective waveform bandwidth is the product of the number of coherently integrated pulses, i.e., $N$ and the frequency step size $\Delta f$. This significantly decreases the ADC sampling rates and only a narrowband, less complex receiver is needed. Consequently, compared to UWB and PC, the SFW radars have long been favored when high-range-resolution profiles (HRRPs) are desired \cite{wehner1995high}. In this paper, we focus on improving aspects of SFW processing.

The synthetic wide bandwidth in traditional linear SFW operation is not without its drawbacks. 
Operationally, it is difficult to maintain the stability of the transmitter and local oscillators for the entire duration of $N$ pulses 
\cite{jankiraman2007design}. 
A serious problem arises when the targets are moving. The SFW receiver is based on \textit{stretch processing} which rapidly range-compresses the radar data using fast Fourier transform (FFT). However, the relative radial velocity between the target and radar induces motion-induced linear and quadratic phase terms \cite{wehner1995high}. The quadratic phase results in \textit{range migration}, i.e., the dispersion of the target response in the HRRP; the severity of the dispersion being proportional to $M$ \cite{wehner1995high}. Due to \textit{range-Doppler coupling}, this class of signals is used only against static targets in applications like GPR \cite{liu2018quantitative} and SAR \cite{perry1999sar}.

A linear sweep across a wide bandwidth makes SFW radars susceptible to co-existing spectral interference. Historically, this was common in SFW-based FOPEN systems \cite{davis2011foliage} that operated at ultra high frequency (UHF) and very high frequency (VHF) dense with radio and television transmissions. 
The FOPEN systems would notch out the interference at the receiver leading to a loss of range resolution. Nowadays, with the advent of several licensed cellular services across multiple IEEE radar bands, 
radar and communications systems are increasingly required to share the spectrum and operate without performance degradation \cite{mishra2019toward,zheng2019radar,ma2019joint,cohen2017spectrum}. In this context, a large bandwidth SFW radar is quite excessive and incompatible with modern spectrum-sharing requirements. 

Finally, a linear SFW radar is not an optimal choice for electronic counter-countermeasures (ECCM) \cite{garmatyuk2002eccm}. Since the carrier frequency variation is always linear, an interceptor can easily track and predict the SF sequence and then implement precise interference to bombard the radar with. Then, the SFW radar suffers from both poor anti-jamming and inability to identify the interception receiver \cite{dudczyk2017method}.

\subsection{Prior Art on SFW Variants}
\label{subsec:prior}
 \begin{table*}
 \centering
 \caption{Comparison of the proposed method with the state-of-the-art}
 \label{tbl:priorcomp}
 \begin{threeparttable}
 \begin{tabular}{l|c|c|c|c|c|c|c|c|c}
 \hline Approach & Carriers\tnote{a} & Sequence\tnote{b} & Pulses\tnote{c} & Targets & Recovery\tnote{d} & Coupling\tnote{e} & Coexistence\tnote{f}  & Interference\tnote{g}  & Delay, Doppler resolution\tnote{h} \\
 \hline \hline SFW \cite{gill1995step,jankiraman2007design} & $M$ & Linear & $N=M$ & Static & Correlation & Strong & No & Strong & $\frac{c}{2M\Delta f}$, N/A \\
 \hline Costas' code \cite{costas1984study} & $M$ & Random (fixed) & $N \geq M$ & Moving & Correlation & Low & No & Strong & $\frac{c}{2M\Delta f}$, $\frac{1}{NT}$  \\
 \hline RSF \cite{axelsson2007analysis} & $M$ & Random & $N \geq M$ & Moving & Correlation & Low & No & Moderate & $\frac{c}{2M\Delta f}$, $\frac{1}{NT}$  \\
 \hline RSF \cite{liu2014decoupled} & $M$ & Random & $N\geq M$ & Static & Sparse & No & No & Moderate & $\propto$ grid size of $M$, N/A \\
 \hline RSF \cite{huang2014cognitive,huang2018analysis} & $M$ & Random & $N\geq M$ & Moving &  Sparse & No & No & Moderate & $\propto$ grid size of $M$, $N$ \\
\hline SSFW \cite{zhang2011sparse,zheng2017lp} & $M$ & Sparse, linear & $N < M$ & Static & Sparse & Strong & Yes & Moderate & $\propto$ grid size of $M$, N/A \\ 
 \hline RaSSteR (this paper) & $M$ & Sparse, random & \rule{0pt}{9pt}$N \gtreqqless M$ & Moving & Sparse & No & Yes & Low & $\propto$ grid sizes independent of $M$, $N$\vspace{2pt}\\
 \hline
 \end{tabular}
 \begin{tablenotes}[para]
 \item[a] Number of unique carrier frequencies available for transmission in the given bandwidth $B$. The SSFW and RaSSteR do not use all frequencies.
 \item[b] Order of transmitting carrier frequencies in a single CPI. 
 \item[c] Number of pulses transmitted in a single CPI. In random waveforms, $N>M$ implies reuse of some of the frequencies in a CPI. In RaSSteR, $N \geq M$ also implies reuse but only from the frequencies within its thinned spectrum. 
 \item[d] Signal recovery method employed by the digital receiver: correlation-based or sparse reconstruction.
 \item[e] Range-Doppler coupling. 
 \item[f] Capability to operate with reduced spectrum in order to share it with other coexisting emitters.
 \item[g] Susceptibility to electronic warfare interference.
 \item[h] In noiseless scenario. The Doppler resolution is not applicable (N/A) to static targets.
 \end{tablenotes}
 \end{threeparttable}\vspace{-12pt}
\end{table*}
Over the past two decades, several variations of SFW radars have been suggested. 
For moving targets, usually motion compensation techniques based on cross-correlation function \cite{prodi2007motion}, improved kinematic modeling \cite{liu2010velocity}, chirp Z-transform \cite{kulpa2006stretch}, and Keystone transform \cite{perry1999sar,zhang2005dim} have been proposed to reduce the adverse impact of Doppler coupling. However, these methods involve a computationally intensive resampling stage in the \textit{slow-time} (CPI) domain and are unable to completely eliminate coupling \cite{pignol2017lagrange}.

It is possible to further mitigate the range-Doppler coupling by randomizing the carrier frequency sequence and follow it with correlation processing of the received signals. The Costas' code \cite{costas1984study} sequence is an example of this processing. Here, unlike SFW, the number of pulses $N$ could exceed the number of available carriers $M$ such that some of the frequencies are reused and each carrier is transmitted at least once. However, the pseudorandom sequence remains fixed across all CPIs in Costas' code and an intelligent interceptor could still track and learn it. This problem is averted in randomized step frequency (RSF) radars 
\cite{axelsson2007analysis} where the sequence is updated in each CPI (Fig.~\ref{fig:waveforms}b). Since the receiver processing in these RSF systems remains conventional correlation-based (with or without motion compensation), the range-Doppler coupling is mitigated but not entirely eliminated. Further, RSF continues to transmit all frequencies within the bandwidth and, thus, the radar remains susceptible to co-existing interference.

Some recent works suggest employing sparse SFW (SSFW) radars \cite{zhang2011sparse,zheng2017lp} to address these drawbacks of RSF processing. The SSFW burst has few ($N < M$) subpulses and some frequencies remain unused during a CPI (Fig.~\ref{fig:waveforms}c). The signal recovery 
relies on the fact that the target scene is sparse facilitating the use of compressed sensing (CS) methods \cite{eldar2012compressed}, which allow recovery of sparse, undersampled signals from random linear measurements. This reduces the duration of each burst and improves the data rate of the radar. Further, interference is directly addressed by skipping the hostile frequency bands. However, the frequency sequence in SSFW is not randomized; only some frequencies in the linear sequence are randomly skipped. SSFW has only been applied to static target scenarios. In this paper, we propose a \textit{ra}ndom, \textit{s}parse \textit{ste}p-frequency \textit{r}adar (RaSSteR) for moving targets. Like SSFW, this new technique has low spectral usage. However, the ECCM performance of RaSSteR is more robust than SSFW because not only does it shuffle vacant bands in each burst but it also randomizes carrier sequence in each CPI (Fig.~\ref{fig:waveforms}d).

Among other prior studies, the closest to our work are CS-based RSF radars \cite{liu2014decoupled,huang2014cognitive,huang2018analysis}. However, the ISAR application in \cite{liu2014decoupled} is for a static scene. A sparse recovery method to estimate the range and Doppler of multiple targets in a coarse range bin is suggested in the RSF framework of \cite{huang2014cognitive}. For the same radar, theoretical performance guarantees are derived in \cite{huang2018analysis}. Contrary to these works, our proposed RaSSteR uses fewer subpulses, is more robust as ECCM, and possesses innate ability to skip the spectrum occupied by other licensed radiators. Table~\ref{tbl:priorcomp} compares our proposed method with these approaches.

\vspace{-10pt}
\subsection{Our Contributions}
\label{subsec:contrib}
Our work has close connections with a rich heritage of research in reduced-rate radar signal processing, see e.g. \cite{mishra2019sub,cohen2018sub} for a recent review of these techniques. For intrapulse PC radars, many prior works address the rate bottleneck in matched filtering by sampling the signals at sub-Nyquist rates and then reconstructing the randomly linear measurements via CS-based recovery. Some prominent approaches include on-grid CS \cite{herman2009high}, off-grid CS \cite{heckel2016superrad}, parametric recovery \cite{bajwa2011identification}, matrix completion \cite{mishra2014compressed,sun2015mimo}, and finite-rate-of-innovation (FRI) models \cite{barilan2014focusing,rudresh2017finite}. In particular, our prior works on sub-Nyquist radars based on FRI are notable for directly addressing the analog sampling \cite{mishra2017sub}, feasible hardware implementations \cite{baransky2014prototype,mishra2019cognitive}, extensions to spatio-temporal-Doppler domains \cite{na2018tendsur}, and robustness to noise \cite{barilan2014focusing} and clutter \cite{eldar2015clutter,li2018sub}. 
Contrary to these studies which are alternatives to intrapulse PC, our work explores interpulse PC in the context of sparse waveforms.

In this work, our main contributions are:\\
\textbf{1) Fewer frequencies and low data rate.} For the same bandwidth, our approach uses fewer transmit frequencies than SFW systems. The traditional low data-rate advantage of SFW over UWB and PC is, therefore, further enhanced by processing fewer PRIs in RaSSteR.\\
\textbf{2) Short CPI.} Less transmit frequencies imply that the burst period in RaSSteR is smaller than both SFW and RSF. This saving in the dwell time without significant information loss leads to higher update rate per scan. However, if further improvement in RaSSteR target detection is desired, $N$ can be matched with SFW/RSF by using some frequencies from the sparse waveform more than once. Hence, in RaSSteR, $N \gtreqqless M$, i.e., $N$ is independent of $M$.\\
\textbf{3) Decoupled range-Doppler.} The lingering problem of errors in SFW measurement of the target's range arising from frequency shift due to target motion is resolved in our formulation. Specifically, the randomization of carrier sequence \cite{cohen2019high} and replacing the standard Fourier-based processing with a CS-based reconstruction completely avoids Doppler shift being reflected as an error in range. Note that RSF with sparse recovery leads to the same result but consumes wider spectrum than RaSSteR. On the other hand, interpolation methods used in classical SFW processing are computationally very intensive.\\
\textbf{4) Guaranteed recovery with less carriers.} We derive strong theoretical performance guarantees for RaSSteR performance. For CS-based RSF systems, \cite{huang2018analysis} shows perfect recovery of $N/2$ targets using all $N=M$ frequencies. We show that RaSSteR perfectly retrieves the same number of targets even when $N<M$.\\
\textbf{5) Delay and Doppler recovery with arbitrary grid size.} In previous formulations \cite{liu2014decoupled,huang2018analysis} with $N\geq M$, the grid size in the discretization of high-resolution delay and Doppler plane was dependent on $M$ and $N$, respectively. However, in RaSSteR where $M$ and $N$ are not dependent on each other, the grid sizes are not restricted by these parameters and could be scaled up to enhance resolution.\\
\textbf{6) Robustness to interference and cognitive spectrum-sharing capability.} By randomly permuting the vacant and used frequencies, RaSSteR provides a stronger defense against electronic attacks. Further, the sparse or \textit{thinned} transmit spectrum of RaSSteR makes it useful as a cognitive radar which must share its operational bandwidth with other communication services, thereby providing a robust solution for coexistence in spectrally crowded environments \cite{mishra2017performance}. We show that, for the same given power budget, RaSSteR outperforms SFW/RSF by focusing all its power in its thinned spectrum. 

Our numerical experiments demonstrate these advantages. We compare the performance of RaSSteR with a conventional SFW \cite{levanon2004radar}, a chirped (PC-based) PDR \cite{peebles1998radar}, and a sub-Nyquist radar \cite{barilan2014focusing} under various target scenarios, signal-to-noise ratios (SNRs), and interference levels. Note that SSFW is used only for static targets and hence, we exclude it from our experiments that comprise more general target scenarios. Our approach exhibits robust, super-resolved, and accurate localization with RaSSteR over state-of-the-art. 

The rest of the paper is organized as follows. In the next section, 
we introduce the system model and formulate the signal recovery in Section~\ref{sec:recovery} and also discuss some of the operational issues related to the interference management. We derive recovery guarantees in Section~\ref{sec:perfanal}, validate our methods through numerical experiments in Section~\ref{sec:numexp}, and conclude in Section~\ref{sec:summ}.

Throughout the paper, we reserve boldface lowercase, boldface uppercase, and calligraphic letters for vectors, matrices, and index sets, respectively. We denote the transpose, conjugate, and Hermitian by $(\cdot)^T$, $(\cdot)^*$, and $(\cdot)^H$, respectively. The identity matrix of size $N\times N$ is $\mathbf{I}_N$. The column-wise Hadamard product is denoted by $\circledcirc$; $||\cdot||_p$ is the $\ell_p$ norm; and $\|\cdot\|_0$ is the number of non-zero elements of the vector. The notation 
$\textrm{det}(\cdot)$ is the determinant, $|\cdot|$ is the cardinality of a set, 
$\mathbb{E}\left[ \cdot \right]$ is the statistical expectation function, and $\mathbb{P}$ denotes the probability. The functions $\text{max}$ and $\text{min}$ output the maximum and minimum values of their arguments, respectively. The function $\text{diag}(\cdot)$ outputs a diagonal matrix with the input vector along its main diagonal. The notation $\lfloor(\cdot)\rfloor$ indicates the greatest integer smaller than or equal to the argument. 

\section{Problem Formulation}
\label{sec:sysmod}
Consider a SFW-based PDR system that transmits a total of $N$ pulses in a CPI with a constant PRT $T$ towards the targets-of-interest. Each of the transmit pulse is nonzero over its support $[0, \tau]$. The duration of each pulse is $T_p$ and the total transmit bandwidth available at the baseband is $B$. 
The frequency of the $n$-th pulse is $f_n \in [f_c, f_c + B]$, $n = 0, 1, \cdots, N-1$, where $f_c$ is the lowest carrier frequency within the sweep $B$. Note that, at RF ranges, $f_c \gg B$. The frequencies $f_n$'s are drawn uniformly at random from the set $\mathcal{F} = \{f_n | f_n = f_c + d_n\Delta f, d_n \in \{0, 1, 2,\cdots,\lfloor B/\Delta f\rfloor=M\}\subset \mathbb{Z} \}$, where $\Delta f$ is the frequency step size. In RaSSteR, $d_n$'s are chosen from a subset of $[0, M]$ so that $M > N$. 
	
The $n$-th transmit pulse is
	\begin{align}
	h(n, t)&=\text{rect}\left(\frac{t-nT}{T_p}\right)e^{\mathrm{j}2\pi (t-nT)f_{n}},
	\end{align} 
where 
	\begin{equation}
	\text{rect}\left(\frac{t-nT}{T_p}\right)=
	\begin{cases} 
	1, \phantom{1}\phantom{1} nT \le t \le nT+T_p,\\
	0, \phantom{1}\phantom{1}\text{otherwise}.
	\end{cases}
	\end{equation} 
All pulses are unit energy waveforms, i.e., $\int_0^{T_p} h(n,t) h^*(n,t) dt=1$ and the total power transmitted by the radar over a single burst is $P_t$.
\vspace{-10pt}
\subsection{Operating Conditions}
Assume that the target scene consists of several non-fluctuating point targets, according to the Swerling-0 target model \cite{skolnik2008radar}. When the radiated wave from the transmitter interacts with moving targets, the amplitude and frequency of the backscattered wave change. The reflected echo from each target at the receive antenna is the attenuated, delayed, and modulated version of the transmit signal. Radar signal processing aims to estimate the following unknown parameters from the received signal: range $r_k$, radial velocity $v_k$; and complex reflectivity $\beta_k$. The target locations are defined with respect to the polar coordinate system of the radar and their range and Doppler are assumed to lie in the unambiguous time-frequency region, i.e., the time delays are no longer than the PRI, and Doppler frequencies are up to the PRF. 

The operating conditions of the radar are as follows:
\begin{description}
  \item [C1] ``Constant delays'': The modulation in frequency arising from a moving target appears as a frequency shift in the received signal. We consider this shift to be small over a CPI so that the delay is approximated to be constant. 
  \item[C2] ``Stop-and-hop Doppler shifts'': At time $n T$ of the transmission of the $n$-th pulse, the target is \textit{stationary} at range $r_k(n)=r_k(0)+\nu_k n T$ and remains so during the flight of the radar pulse. The target then ``hops'' to the position $r_k(n+1)=r_k(0)+\nu_k (n+1) T$. It then stays there until the time $(n+1) T$ when the $(n+1)$-th pulse is transmitted and so forth. Hence, at the times of pulse transmission, the target is at exactly the same locations it would have occupied as if it was moving at a constant velocity, i.e., $r_k(t)=r_k(0)+\nu_k t$. However, for individual pulses, the target is modeled as if the $n$-th pulse is reflected from a stationary target at range $r_k(m)$. In this model, there is no actual Doppler shift because the target is stationary during each pulse transmission; rather there is only a pulse-to-pulse variation in the target's range. This assumption gives rise to the notion of a \textit{coarse} range that corresponds to the location during the first pulse transit and a \textit{fine} or high-resolution range (HRR) arising from the inter-pulse range variation during the CPI.
  \item[C3] ``Small acceleration'': The acceleration in the movement of the target is small so that the targets do not move beyond the coarse range cell over the CPI.
  \item[C4] ``Constant reflectivities'': The targets are in the far ranges so that the complex reflectivities are constant during the observation time $t \in [0,(N-1)T]$.
  \item[C5] ``Unambiguous HRRP'': In order to avoid the ``ghost images'' within the HRR bin, the unambiguous scope of HRRP defined by $c /(2 \Delta f)$ should be larger than the scope of a coarse-range bin $c T_{p} / 2$. This implies $\Delta f<1 / T_{p}$.
\end{description}
    \vspace{-10pt}
    \subsection{Receive Signal}
	With these assumptions, the received echo of the $n$-th pulse from the $k$-th target is
	\begin{align}
	g_{k}(n, t)\approx\beta_kh\left(n, t-\frac{2r_k(t)}{c}\right),
	\end{align} 
	where $r_k(t)$ is the range of the target at instant $t$ and the approximation arises by ignoring the acceleration. In the stop-and-hop model, the target's constant Doppler is $\nu_k(t)$ so that $r_k(t) = r_k(0) + \nu_k t$. 
	
	The $n$-th echo is sampled at the rate $1/T_p$ in \textit{fast-time} $t_s(n, l_r) = nT + l_rT_p$, where $l_r = 1, \cdots, L_r$, $L_r = \lfloor T/T_p\rfloor$. The maximum unambiguous range is $R_{max} = cT/2$. The sampled, discrete-time received signal for the $n$-th pulse is
	\begin{align}
	g_{k}[n, t_s(n,l_r)] = \beta_k e^{\mathrm{j}2\pi \left(-2\left(\frac{R_k+\nu_knT}{c} \right) \right) f_{n}},
	\end{align} 
	where
	\begin{align}
	\label{eq:coarsecell}
	R_k = r_k(0)-\frac{l_rT_pc}{2},
	\end{align}
	is the HRR of the $k$-th target.
	
	Assuming there are $K$ targets in the $l_r$-th coarse range cell, and negligible mutual coupling of echoes therein, the sampled received signal is the superposition of echoes from all targets, i.e.,
	\begin{align}
	g[n, t_s(n,l_r)] &= \sum_{k=1}^{K}g_k[n, t_s(n,l_r)]\nonumber\\
	&= \sum_{k=1}^{K}\beta_k e^{\mathrm{j}2\pi \left(-2\left(\frac{R_k+\nu_knT}{c} \right) \right) f_{n}}.\label{eq:rxsig3}
	\end{align} 
	Using $f_n = f_c +d_n \Delta f$ in \eqref{eq:rxsig3} above and ignoring the cross term because $f_c \gg \Delta f$ yields the approximation
	\begin{align}
	g[n, t_s(n,l_r)] & \approx \sum_{k=1}^{K}\beta_k e^{-\mathrm{j}\frac{4\pi}{c}f_c R_k} e^{-\mathrm{j}\frac{4\pi}{c} d_n \Delta fR_k} e^{-\mathrm{j}\frac{4\pi}{c}f_c \nu_k nT}\\
	& =\sum_{k=1}^{K} \gamma_k e^{-\mathrm{j} 2\pi \frac{R_k}{R_u}d_n} e^{-\mathrm{j} 2\pi \frac{\nu_k}{\nu_u}n},
	\end{align} 
	where 
	\begin{align}
	\label{eq:gamma}
	\gamma_k =\beta_k e^{-\mathrm{j}\frac{4\pi}{c}f_c R_k} =\beta_k e^{-\mathrm{j}\frac{4\pi}{c}f_c \left(r_k(0)-\frac{l_rT_pc}{2}\right)}.
	\end{align}
	
	The quantities $R_{u}=\frac{c}{2\Delta f}$ and $\nu_{u}=\frac{c}{2 f_c T}$ denote the maximum unambiguous high-resolution range and unambiguous velocity, respectively. We assume all targets satisfy the inequalities 
	\begin{align}
	-\frac{R_u}{2}<R_k < \frac{R_u}{2}, \quad \text{and} \quad 0\leq \nu_k < \nu_u.
	\end{align}
	For uniqueness of recovery, $R_u$ should be greater then the coarse range, that is, $R_u \geq \frac{T_p c}{2}$. This implies that $\Delta f \leq \frac{1}{T_p}$.
	
	Assume that the targets lie on a grid in the high-resolution range and Doppler plane with grid resolutions $\Delta R=\frac{R_u}{P}$ and $\Delta \nu = \frac{\nu_u}{Q}$, respectively, where $P, Q \in \mathbb{N}$. For each target, we have $R_k=n_k \Delta R $ and $\nu_k=m_k \Delta \nu$, where $n_k \in \{0, 1,\cdots, P-1\}$ and $m_k\in \{0, 1,\cdots, Q-1\}$. After this discretization, the measurements are 
	\begin{align}
	g[n, t_s(n,l_r)] =\sum_{k=1}^{K} \gamma_k e^{\mathrm{j} p_k d_n} e^{\mathrm{j} q_kn},
	\label{eq:mes1}
	\end{align}
	where $p_k=-\frac{2\pi}{P}n_k$ and $q_k=-\frac{2\pi}{Q}m_k$. Our goal is to recover $n_k$, $m_k$, and $\gamma_k$ from $N$ pulses which need not be same as $M$ (as is the case with SFW). 
    \vspace{-10pt}
    \section{Target Recovery}
    \label{sec:recovery}
    In a traditional SFW system, the received echo from a particular range is coherently detected and the phase of the resulting signal is then stored for all pulses. For a sequence of uniformly-spaced frequency steps, the inter-pulse phase shift of $\Delta \phi$ is a function of the frequency difference as $\Delta \phi=\frac{4 \pi \Delta f r_k(0) }{c}$ from which the range of the static target is easily extracted. In case multiple targets occur in the same range bin, each generates a unique frequency that is extracted from the time domain signal using FFT. If all the targets spanning the full unambiguous range fall in several coarse range bins, then each bin needs to be processed by FFT. 

    When the targets are moving, the SFW received signal corresponding to \eqref{eq:mes1} has exponential terms with $d_n$ replaced by the common variable $n$ leading to coupling of range and Doppler information. In this case, distinguishing these two parameters from the phase shift $\Delta \phi$ alone is very difficult and only approximate retrieval is possible through compensation techniques which must be applied across the entire HRRP leading to significant computational load \cite{prodi2007motion,kulpa2006stretch,perry1999sar,zhang2005dim}. When the frequencies are randomized, the appearance of the variables $d_n$ in \eqref{eq:mes1} is key to decoupling as explained below.
    \vspace{-10pt}
    \subsection{Sparse Reconstruction}
    \label{subsec:reconst}
	For a given $l_r$-th coarse range bin that contains $K$ targets, we collect echoes for $N$ pulses and denote the $N\times 1$ measurement vector
	\begin{align}
	\mathbf{\tilde{y}} = \begin{bmatrix} g[0, t_s(0,l_r)] \\ g[1, t_s(1,l_r)] \\ \vdots \\ g[N-1, t_s(N-1,l_r)] \end{bmatrix} = \begin{bmatrix} \sum_{k=1}^{K} \gamma_k e^{\mathrm{j} p_k d_0} \\ \sum_{k=1}^{K}\gamma_k e^{\mathrm{j} p_k d_1} e^{\mathrm{j} q_k} \\ \vdots \\ \sum_{k=1}^{K}\gamma_k e^{\mathrm{j} p_k d_{N-1}} e^{\mathrm{j} q_k(N-1)} \end{bmatrix},
	\end{align}
	where, from \eqref{eq:gamma}, $\gamma_k$ is dependent on the index $l_r$ of the fixed coarse range cell. We cast these discretized measurements as a sparse localization model
	\begin{widetext}
	\begin{align}
	\centering
	\mathbf{\tilde{y}}=
	&\left[
        \begin{array}{cccc|ccc|cc}
        1 & 1 & \cdots & 1 & e^{-\mathrm{j}2\pi\frac{d_0}{P}} 
        & \cdots & e^{-\mathrm{j}2\pi\frac{d_0}{P}}  & \cdots & e^{-\mathrm{j}2\pi\frac{(P-1)d_0}{P}}\\
        1 & e^{-\mathrm{j}2\pi\frac{1}{Q}} & \cdots & e^{-\mathrm{j}2\pi\frac{(Q-1)}{Q}} & e^{-\mathrm{j}2\pi\frac{d_1}{P}} 
        & \cdots & e^{-\mathrm{j}2\pi\left(\frac{d_1}{P}+\frac{(Q-1)}{Q}\right)}  & \cdots & e^{-\mathrm{j}2\pi\left(\frac{(P-1)d_{1}}{P} + \frac{(Q-1)}{Q}\right)}\\ 
	    \vdots & \vdots & \ddots & \vdots & \vdots 
	    & \ddots & \vdots & \ddots & \vdots \\ 
	    1 & e^{-\mathrm{j}2\pi\frac{N-1}{Q}} & \cdots & e^{-\mathrm{j}2\pi\frac{(N-1)(Q-1)}{Q}} & e^{-\mathrm{j}2\pi\frac{d_{N-1}}{P}} 
	    & \cdots & e^{-\mathrm{j}2\pi\left(\frac{d_{N-1}}{P} + \frac{(N-1)(Q-1)}{Q}\right)} & \cdots & e^{-\mathrm{j}2\pi\left(\frac{(P-1)d_{N-1}}{P} + \frac{(N-1)(Q-1)}{Q}\right)}
        \end{array}
    \right]\mathbf{x}\nonumber\\
	&=\mathbf{A}\mathbf{x},
	\label{eq:mes2}
	\end{align}
	\end{widetext}
	\noindent where $\mathbf{A}\in \mathbb{C}^{N \times PQ}$ is a measurement matrix and $\mathbf{x}\in \mathbb{C}^{PQ}$ is a $K$-sparse vector whose nonzero entries are complex reflectivities at indices $u_k=n_k+m_kQ$ for $k=1,2,\cdots,K$. Note that the measurement matrix in \cite{huang2018analysis} is $M\times PQ$, $M>N$ and $P$ and $Q$ dependent on $M$ and $N$, respectively.
	
	The measurement matrix $\mathbf{A}$ in \eqref{eq:mes2} is decomposed as
	\begin{widetext}
	\begin{align}
	\mathbf{A}&= \begin{bmatrix}
	1 & e^{-\mathrm{j}2\pi\frac{d_0}{P}} & \cdots & e^{-\mathrm{j}2\pi\frac{(P-1)d_0}{P}}\\ 
	1 & e^{-\mathrm{j}2\pi\frac{d_1}{P}} & \cdots & e^{-\mathrm{j}2\pi\frac{(P-1)d_1}{P}}\\ 
	\vdots & \vdots & \ddots & \vdots\\ 
	1 & e^{-\mathrm{j}2\pi\frac{d_{N-1}}{P}} & \cdots & e^{-\mathrm{j}2\pi\frac{(P-1)d_{N-1}}{P}}
	\end{bmatrix}\circledcirc \begin{bmatrix}
	1 & 1 & \cdots & 1\\ 
	1 & e^{-\mathrm{j}2\pi\frac{1}{Q}} & \cdots & e^{-\mathrm{j}2\pi\frac{(Q-1)}{Q}}\\ 
	\vdots & \vdots & \ddots & \vdots\\ 
	1 & e^{-\mathrm{j}2\pi\frac{N-1}{Q}} & \cdots & e^{-\mathrm{j}2\pi\frac{(N-1)(Q-1)}{Q}}
	\end{bmatrix}\nonumber\\
	&=\mathbf{R} \circledcirc \mathbf{D} 
	= [\text{diag}(\mathbf{r}_1) \mathbf{D},\,\, \text{diag}(\mathbf{r}_2) \mathbf{D},\,\,\cdots , \text{diag}(\mathbf{r}_P) \mathbf{D} ],
	\label{eq:amat}
	\end{align}
	\end{widetext}
	where the ${N\times P}$ matrix $\mathbf{R}$ represents the high-resolution range dictionary whose $(n,p)$-th element is $\mathbf{R}[n,p]=e^{-\mathrm{j}\frac{2\pi}{P}p d_n}$; the ${N\times Q}$ matrix $\mathbf{D}$ is the Doppler dictionary with $(n,q)$-th element $\mathbf{D}[n,q]=e^{-\mathrm{j}\frac{2\pi}{Q}qn}$; and $\mathbf{r}_p$ denotes the $p$-th column of $\mathbf{R}$. Note that each column of the matrix $\mathbf{A}$ is indexed by the pair $(p, q)$ where $p\in \{0, 1, \cdots, P-1\}$ and $q \in \{ 0, 1, \cdots, Q-1\}$.
	
	Assume the measurements $\mathbf{\tilde{y}}$ are contaminated with additive white Gaussian noise (AWGN) at the receiver so that the corrupted data are
    \begin{align}
    \mathbf{y}=\mathbf{A} \mathbf{x} + \mathbf{z},\label{eq:mes3}
    \end{align}
    where $\mathbf{z}$ is the noise vector drawn from the complex normal distribution $\mathcal{C} \mathcal{N}\left (0, \sigma^{2} \mathbf{I}_{N}\right)$, where $\sigma^{2}$ is the noise variance. To recover $\mathbf{x}$, consider a sparse recovery problem
    \begin{flalign}
    \label{eq:opt}
	& \underset{\mathbf{x}}{\text{minimize}}\phantom{1}\left\Vert \mathbf{x}\right\Vert _{0}\nonumber\\
	& \text{subject to}\phantom{1} \left\Vert \mathbf{y}-\mathbf{A} \mathbf{x}\right\Vert_2 \le \eta,
\end{flalign}\normalsize
    where $\eta$ is some positive constant dependent on the noise variance. This problem can be solved using one of the several compressed sensing recovery algorithms such as via $\ell_1$ minimization and greedy algorithms like orthogonal matching pursuit (OMP) \cite{eldar2012compressed}. In the numerical experiments of Section~\ref{sec:numexp}, we adopt OMP for sparse recovery in RaSSteR.
    
    The recovered $\mathbf{x}$ has non-zero entries at indices $u_k$. The high-resolution range and Doppler are given by the index set\par\noindent\small
    \begin{align}
    \label{eq:suppset}
    \Lambda = \{(n_k,m_k)|n_k = u_k \bmod Q, m_k = \lfloor u_k/Q \rfloor, k=1,\cdots,K\}.
    \end{align}\normalsize
    Then, for $k=1,\cdots,K$, the targets' ranges, Doppler velocities and reflectivities are
    \begin{align}
    R_k = n_k \Delta R, \nu_k = m_k \Delta \nu,\; \textrm{and} \; \beta_k = \gamma_k e^{\mathrm{j}\frac{4\pi}{c}f_c R_k},
    \end{align}
    respectively. It follows from \eqref{eq:amat} that the range and Doppler phases are separable or uncoupled. Moreover, our non-reliance on FFT-based processing eliminates the difficulty in estimating range and Doppler independent of each other.

    \subsection{Interference and Cognitive SFW}
    \label{subsec:cogsfw}
	The carrier frequency selection is also decided by the presence of interference. For instance, if a strong interference lies in the frequency band $[f_c+M_1 \Delta f, f_c+M_2 \Delta f ]$ for $0 < M_1 < M_2 < M$, then the radar chooses  $d_n$'s randomly from $[0, M_1+1]\cup [M_2+1, M]$ to avoid transmitting in the hostile frequencies. Assuming the set of $L$ such disjoint subbands is
	\begin{align}
	\mathcal{I} = \cup_{l=1}^L \mathcal{I}_l \subset \mathbb{R},
	\label{eq:interval}
	\end{align}
	we draw $d_n$ uniformly at random from available frequencies in the set $\mathcal{I}$. Here, the height of the (uniform) probability density of variable $d_n$ is same across all the intervals. Specifically, we assume that the density in each interval is $\frac{1}{|\mathcal{I}|}$ where
    \begin{align}
    |\mathcal{I}| = \cup_{l=1}^L |\mathcal{I}_l| <M.
    \label{eq:density}
    \end{align}
    
    Denote the transmit pulses of SFW and RaSSteR by $h_{\textrm{S}}(n,t)$ and $h_{\textrm{R}}(n,t)$, respectively. If both SFW and RaSSteR systems are allowed the same total burst power, i.e., \par\noindent\small
    \begin{align}
        \sum_{n=1}^M \int_0^{T_p} h_{\textrm{S}}(n,t) h_{\textrm{S}}^*(n,t) dt= \sum_{ {\substack{{n = 1}\\{d_n \in \mathcal{I} }}} }^N \int_0^{T_p} h_{\textrm{R}}(n,t) h_{\textrm{R}}^*(n,t) dt= P_t,
    \end{align}\normalsize
    then RaSSteR, which transmits fewer $N<M$ pulses, would have higher per pulse power than SFW. This leads to an increase in SNR at the RaSSteR receiver, thereby providing better hit rates than SFW (as well as RSF) systems with the same energy budget. We employ this focusing of all the available power in the RaSSteR pulses as a deterrent in spectral interference situations. This technique mirrors the cognitive operation that was earlier proposed for sub-Nyquist processing of intrapulse PC radars \cite{mishra2017performance,mishra2019cognitive}. 

    \subsection{Clustered Targets}
    \label{subsec:clust}
    The HRRPs in conventional SFW radars are often required to distinguish objects that occur in dense clusters. For example, military FOPEN systems are often used to identify groups of human intruders localized under a thick forest canopy \cite{davis2011foliage}. Similarly, an ISAR system could be interested in imaging a formation of multiple aircraft \cite{soumekh1999synthetic}. Since synthesizing a full HRRP using motion-compensated-SFW in these applications entails processing hundreds of range bins, the detection of a few clusters of targets easily becomes computationally very intensive. 

    On the other hand, coarse-bin-based RaSSteR processing implies that, without applying the motion compensation across the entire coarse range, it is possible to retrieve HRRP within each coarse bin. This technique is, therefore, very efficient in practice for detecting each object in a dense cluster. Based on a predetermined probability of false alarm $P_{\textrm{fa}}$, only those coarse bins are processed for HRRP whose SNR exceeds a certain threshold $\gamma$, which is obtained from classical detection theory considerations \cite{poor1994introduction}. In particular, the generalized likelihood ratio test (GLRT) applied to each coarse range cell posits the alternative and null hypotheses to define the presence or absence of a clustered target, respectively. If $Q_{\chi_2^2(\rho)}$ denotes the right-tail probability function of the chi-square distribution with $2$ degrees of freedom, $\sigma^2$ is the noise variance, and $\rho=P_t/(\sigma^2|\mathcal{I}|)$ is the SNR, then $\gamma = Q^{-1}_{\chi_2^2(\rho)}(1-(1-P_{\text{fa}})^{|I|})$.
    
    In this case, fewer coarse range cells are investigated for target swarms and applying a greedy method such as OMP would require a search over a smaller delay-Doppler grid. On the other hand, reduced-rate processing such as in the FRI-model-based sub-Nyquist radar \cite{barilan2014focusing} would require OMP search over a much larger grid comprising all coarse bins.

    \subsection{Frequency Reuse}
    \label{subsec:reuse}
    The seminal work in \cite{levanon2002stepped} proposed repetitions of frequencies within a CPI as an effective means to reduce the range sidelobes of a single-use SFW sweep. A coherent processing of $N > M$ pulses also enhances the detection rate. The framework in \cite{levanon2002stepped} suggested many different sequences for repetitions such as up-down, linear, nonlinear, and selective repetition of only a few carriers. Our numerical experiments in Section~\ref{sec:numexp} show that an analogous improvement in the RaSSteR detection performance is indeed possible if the number of pulses $N$ is larger than the number of available carriers $|\mathcal{I}|$ so that some frequencies from the sparse spectrum are reused. In this case, $d_n$ is drawn from $\mathcal{I}$ with replacement. In the absence of interference, frequency reuse in RaSSteR leads to only a limited improvement in detection than a similar reuse in RSF. However, this performance is significantly enhanced when signal-to-interference-ratio (SIR) is decreased.
    
\section{Performance Analyses}
\label{sec:perfanal}
We now develop guarantees for target recovery in a RaSSteR system. In particular, we discuss conditions on the number of pulses required to retrieve information from a given number of targets based on the properties of the measurement matrix $\mathbf{A}$. The guarantees derived in \cite{huang2018analysis} for an RSF radar require all $M$ carriers. In contrast, we prove that perfect recovery is also possible with $N<M$ carriers.

	\subsection{Spark}
	\label{subsec:spark}
	The conditions for $\ell_0$ minimization are based on $\text{Spark}(\mathbf{A})$, the spark of the matrix $\mathbf{A}$, which is the smallest number of linearly dependent columns of a matrix. This strategy has been employed in FRI-based sub-Nyquist radars \cite{cohen2016summer} and optical imaging \cite{mulleti2016fri}.
 
	\begin{theorem}
		\label{thm:spark}
		Consider an $N \times PQ$ measurement matrix $\mathbf{A}$ as in \eqref{eq:amat}. Assume that $d_n$'s are independent and identically distributed uniformly over a set of intervals as in \eqref{eq:interval}. The matrix has spark $N+1$ with probability one.   
	\end{theorem}
	\begin{IEEEproof}
		From the principle of mathematical induction, we show that for $N_1 \leq N$, any $N_1 \times N_1$ submatrix of $\mathbf{A}$ is full rank. Denote an arbitrary  $2 \times 2$ submatrix of $\mathbf{A}$ by
		\begin{align}
		\mathbf{A}_2 = 
		\begin{pmatrix}
		e^{\mathrm{j}\frac{2\pi p_1}{P}d_{n_1}+\mathrm{j}\frac{2\pi q_1}{Q}n_1}  & e^{\mathrm{j}\frac{2\pi p_2}{P}d_{n_1}+\mathrm{j}\frac{2\pi q_2}{Q}n_1} \\
		e^{\mathrm{j}\frac{2\pi p_1}{P}d_{n_2}+\mathrm{j}\frac{2\pi q_1}{Q}n_2}  & e^{\mathrm{j}\frac{2\pi p_2}{P}d_{n_2}+\mathrm{j}\frac{2\pi q_2}{Q}n_2}
		\end{pmatrix}.
		\label{eq:2by2}
		\end{align}
		The submatrix is constructed by considering rows $n_1$ and $n_2$ of $\mathbf{A}$. The columns are denoted by the pairs $(p_1, q_1)$ and $(p_2, q_2)$. The matrix $\mathbf{A}_2$ is not full rank if $\text{det}(\mathbf{A}_2)=0$, that is, whenever
		\begin{flalign}
		&e^{\mathrm{j}\frac{2\pi p_1}{P}d_{n_1}+\mathrm{j}\frac{2\pi q_1}{Q}n_1} e^{\mathrm{j}\frac{2\pi p_2}{P}d_{n_2}+\mathrm{j}\frac{2\pi q_2}{Q}n_2}\nonumber\\ 
		&= e^{\mathrm{j}\frac{2\pi p_2}{P}d_{n_1}+\mathrm{j}\frac{2\pi q_2}{Q}n_1} e^{\mathrm{j}\frac{2\pi p_1}{P}d_{n_2}+\mathrm{j}\frac{2\pi q_1}{Q}n_2}, \nonumber
		\end{flalign}
		or,
		\begin{flalign}
		&\frac{p_1}{P}d_{n_1}+\frac{q_1}{Q}n_1+\frac{ p_2}{P}d_{n_2}+\frac{q_2}{Q}n_2 \nonumber\\
		&= \frac{ p_2}{P}d_{n_1}+\frac{ q_2}{Q}n_1+\frac{p_1}{P}d_{n_2}+\frac{q_1}{Q}n_2 +m, 
		\label{eq:det0}
		\end{flalign}	 
		for some $m \in \mathbb{Z}$. For each $m \in \mathbb{Z}$ the set
		\begin{align}
		\mathcal{S}_m = \{(d_{n_1}, d_{n_2}) \in \mathbb{R}^2, \,\, \text{s.t.} \,\,\eqref{eq:det0} \,\, \text{holds} \} 
		\end{align}
		represents a line in $\mathbb{R}^2$ and has zero Lebesgue measure. The set of all $(d_{n_1}, d_{n_2}) \in \mathbb{R}^2$ such that $\text{det}(\mathbf{A}_2)=0$ is given as $\displaystyle \mathcal{S} = \cup_{m \in \mathbb{Z}} \mathcal{S}_m$. The set $\mathcal{S}$ has Lebesgue measure zero as it is a countable union of sets of zero measure. Hence, the Lebesgue measure of the set $\{d_{n}\}_{n=1}^N$ such that any $2 \times 2$ submatrix of $\mathbf{A}$ has rank one is zero. 
		
		Next, denote the Lebesgue measure of the set $\{d_{n}\}_{n=1}^N$ such that all $(N-1) \times (N-1)$ submatrices of $\mathbf{A}$ are non-invertible is zero. Consider a matrix $\mathbf{A}_N$ that consists of a set of arbitrarily chosen $N$ columns of $\mathbf{A}$. Then,
		\begin{align}
		\text{det}(\mathbf{A}_N) = \sum_{n=1}^{N} c_n \, e^{\mathrm{j}\frac{2\pi p_{n}}{P}d_{N}} e^{\mathrm{j}\frac{2\pi q_{n}}{Q}d_{n}},
		\label{eq:detn}
		\end{align}
	where $c_n$ denotes $n$-th cofactor of the matrix $\mathbf{A}_N$. Since $c_n$ is equal (up to a sign factor) to the determinants of $(N-1) \times (N-1)$ submatrix of $\mathbf{A}_N$, the measure of the set $\{d_{n}\}_{n=1}^{N-1}$, such that $\{c_n\}_{n=1}^{N}$ are zero, is zero. The set $\{d_N \in \mathbb{R}: \text{det} (\mathbf{A}_N) = 0\}$ denotes the roots of the analytic function $\text{det}(\mathbf{A})_N$ and it has zero Lebesgue measure. Hence, the measure of the set $\{d_{n}\}_{n=1}^{N}$ such that any $N \times N$ submatrix of $\mathbf{A}$ has low rank is zero. Since each $d_{n}$ is uniformly distributed over an interval on $\mathbb{R}$, any $N \times N$ submatrix of $\mathbf{A}$ has full rank with probability one. Hence, $\text{Spark}(\mathbf{A}) = N+1$ with probability 1.
	\end{IEEEproof}
	
	Theorem~\ref{thm:spark} implies that with the transmission of $N$ pulses, $N/2$ targets can be uniquely identified. Since $N<M$ in RaSSteR, identical recovery performance is achieved with a sparse transmission when compared to RSF \cite{huang2018analysis}. Full spark property of the measurement matrix implies that there are no grating lobes (i.e., no aliasing or ghost images). Specifically, it implies that no two rows of the measurement matrix $\mathbf{A}$ are linearly dependent. This also results in a choice of the grid sizes $P$ and $Q$ independent of the number of pulses $N$ and the number of feasible carrier frequencies $M$. Specifically, we need not choose $P$ and $Q$ such that $M \geq P$ and $Q=N$ as proposed in \cite{huang2018analysis}. Our construction here is not asymptotic in $N$ as in the similar guarantees derived in \cite{huang2018analysis}.
	
	\subsection{Mutual Coherence}
	\label{subsec:mut_coh}
	The recovery guarantees for $\ell_1$ minimization are based on the restricted isometry property (RIP) and mutual coherence of the measurement matrices \cite{eldar2012compressed}. For the measurement matrix $\mathbf{A}$, let $\mathbf{a}_m$ denote its $m$-th column. The coherence of $\mathbf{A}$ is
	\begin{align}
	\mu(\mathbf{A}) = \underset{m \neq \ell}{\text{max}} \quad \frac{\left|\mathbf{a}_m^{\mathrm{H}}\mathbf{a}_\ell \right|}{ \|\mathbf{a}_m\|_2 \|\mathbf{a}_\ell\|_2}.
	\label{eq:mip}
	\end{align}
For matrix $\mathbf{A}$ in \eqref{eq:amat}, we have that $\|\mathbf{a}_m\|_2=\sqrt{N}$. The inner product in \eqref{eq:mip} is given as
\begin{align}
\mathbf{a}_m^{\mathrm{H}}\mathbf{a}_\ell =\sum_{n=1}^N e^{\mathrm{j} \frac{2\pi}{P} (p_m-p_\ell)d_n} e^{\mathrm{j} \frac{2\pi}{Q} (q_m-q_\ell)n},
\end{align}
where the $m$-th and $\ell$-th columns of $\mathbf{A}$ are indexed by ordered pairs $(p_m, q_m)$ and $(p_\ell, q_\ell)$, respectively. Since $|\mathbf{a}_m^{\mathrm{H}}\mathbf{a}_\ell|=|\mathbf{a}_\ell^{\mathrm{H}}\mathbf{a}_m|$, $\mu(\mathbf{A})$ is given as
\begin{align}
\mu(\mathbf{A}) = \max_{\substack{p \in \llbracket 1, P-1\rrbracket, \\ q \in \llbracket 1, Q-1\rrbracket}}\quad \left|\frac{1}{N}\sum_{n=1}^N e^{\mathrm{j} \frac{2\pi}{P} p d_n} e^{\mathrm{j} \frac{2\pi}{Q}q n} \right|.
\label{eq:mip1}
\end{align}

It has been shown \cite{eldar2012compressed} that the $\ell_1$-relaxation of the optimization problem in \eqref{eq:opt} can be efficiently solved provided that $\mu(\mathbf{A})$ is bounded by $\frac{1}{2K-1}$, that is, $\mu(\mathbf{A})<\frac{1}{2K-1}$. Since $\mu(\mathbf{A})$ is a random variable, we show that it satisfies the inequality with high probability provided that the number of targets are restricted. Since the summation in \eqref{eq:mip1} is divided by $N$, not by $\sqrt{N}$, one can not apply either the classical central limit theorem or its Lyapunov's version to derive the distribution of $\mu(\mathbf{A})$ as $N \rightarrow \infty$.  
Recall the bounded differences inequality (BDI) \cite{wainwright2019high} that generalizes Hoeffding’s inequality.
\begin{lemma}{\textup{(Bounded Differences Inequality)}}
\label{thm:bdi}
Consider a set $\mathcal{I}$ and a function $f: \mathcal{I}^N \rightarrow \mathbb{R}$. Let $d_1, d_2, \cdots, d_N$ are set of independent random variables on the set $\mathcal{I}$. If there exist positive real numbers $c_1, c_2,\cdots, c_N$ such that for $n=1, 2,\cdots, N$
\begin{align}
    \underset{d_1,\cdots, d_N, d_n^{\prime} \in \mathcal{I}}{\sup} |f(d_1,..,d_n,.., d_N)-f(d_1,.., d_n^{\prime},.., d_N)| \leq c_n \nonumber
\end{align}
then for all $\lambda>0$
\begin{align}
\mathbb{P} \left(\left| f - \mathbb{E}\left[f\right] \right| \geq \lambda\right) \leq e^{-\frac{\lambda^2}{\sum_{n=1}^N c_n^2}}. \label{eq:BDI} \nonumber
\end{align}
\end{lemma}
\begin{IEEEproof}
We refer the reader to \cite{wainwright2019high} for the proof.
\end{IEEEproof}

We use the BDI to derive a guarantee based on mutual coherence of the matrix $\mathbf{A}$ in the following theorem.
\begin{theorem}
\label{thm:mp}
	Let $d_n$'s be distributed as per \eqref{eq:interval} with parameter $|\mathcal{I}|$ as in \eqref{eq:density}. The coherence of the matrix $\mathbf{A}$ in \eqref{eq:amat} is upper bounded by $\frac{1}{2K-1}$ with probability $1-\delta$ if 
	\begin{align}
	K \leq \frac{1}{2} \sqrt{\frac{N}{2(\log 2(P-1)(Q-1) -\log \delta)}}+\frac{1}{2},
	\end{align} 
	and 
	\begin{align}
	K \leq \frac{\pi |\mathcal{I}|}{2}+\frac{1}{2}.
	\end{align}
\end{theorem}
\begin{IEEEproof}
For a given ordered pair $(p, q)$, define the following functions of the random variables $\{d_n\}_{n=1}^N$:
\begin{align}
f_{(p,q)}^r(d_1, d_2, \cdots, d_N) &= \frac{1}{N}\sum_{n=1}^N \sin\left(\frac{2\pi}{P} p d_n+ \frac{2\pi}{Q}q n\right),
\label{eq:fr} \\
f_{(p,q)}^i(d_1, d_2, \cdots, d_N) &= \frac{1}{N}\sum_{n=1}^N \cos\left(\frac{2\pi}{P} p d_n+ \frac{2\pi}{Q}q n\right)
\label{eq:fi}.
\end{align}
For $n =1, 2, \cdots, N$ we have
\begin{flalign}
&\sup_{d_1, \cdots, d_N, d^{\prime}_n} \left| f_{(p,q)}^r(d_1,\cdots, d_{n-1}, d_n, d_{n+1}, d_N)\right.\nonumber\\
&\;\;\;\;\;\;\;\;\;\;\;\;\;\;\;\;\;\;\;\;\;\;\;\;\;\;\;\;\;\;\;\;\;\;\;\;\left. -f_{(p,q)}^r(d_1,\cdots, d_{n-1}, d_n^{\prime}, d_{n+1}, d_N)    \right| \nonumber \\ 
&= \sup_{d_1, \cdots, d_N, d^{\prime}_n} \frac{1}{N}\left| \sin\left(\frac{2\pi}{P} p d_n+ \frac{2\pi}{Q}q n\right)\right.\nonumber\\
&\;\;\;\;\;\;\;\;\;\;\;\;\;\;\;\;\;\;\;\;\;\;\;\;\;\;\;\;\;\;\;\;\;\;\;\;\left.-\sin\left(\frac{2\pi}{P} p d_n^{\prime}+ \frac{2\pi}{Q}q n\right)   \right| \leq \frac{2}{N}.
\end{flalign}
Similarly,  
\begin{align}
&\sup_{d_1, \cdots, d_N, d^{\prime}_n} \left| f_{(p,q)}^i(d_1,.. d_{n-1}, d_n, d_{n+1}, d_N)\right.\nonumber\\
&\;\;\;\;\;\;\;\;\;\;\;\;\;\;\;\;\;\;\left.-f_{(p,q)}^i(d_1,.. d_{n-1}, d_n^{\prime}, d_{n+1}, d_N)    \right| \leq \frac{2}{N}.
\end{align}
As the random variables $\{d_n\}_{n=1}^N$ are independent, the BDI (Lemma~\ref{thm:bdi}) yields
\begin{align}
\mathbb{P} \left(\left| f_{(p,q)}^r - \mathbb{E}\left[f_{(p,q)}^r\right] \right| \geq \lambda\right) \leq e^{-\frac{N\lambda^2}{2}}, \label{eq:BEr}\\
\mathbb{P} \left( \left|f_{(p,q)}^i - \mathbb{E}\left[f_{(p,q)}^i\right] \right|  \geq \lambda\right) \leq e^{-\frac{N\lambda^2}{2}}, \label{eq:BEi}
\end{align}
for all $\lambda >0$. 

In general, $\mathbb{E}\left[f_{(p,q)}^r\right]=0$ and $\mathbb{E}\left[f_{(p,q)}^i\right]=0$ do not hold. Assume that for a given distribution of $\{d_n\}_{n=1}^N$, $\exists$ $\epsilon >0$ such that
\begin{align}
 |\mathbb{E}\left[f_{(p,q)}^r\right]| \leq \epsilon \quad \text{and} \quad  |\mathbb{E}\left[f_{(p,q)}^i\right]| \leq \epsilon \label{eq:mean_condition},
\end{align}
for all $p \in \llbracket 1, P-1 \rrbracket$ and $q \in \llbracket 1, Q-1 \rrbracket$. Note that, in the RSF analyses of \cite{huang2018analysis}, the corresponding expectations in \eqref{eq:mean_condition} vanish. Next, consider the event $\{|f^r_{(p,q)}|\geq \epsilon\}$. Since $|\mathbb{E}\left[f_{(p,q)}^r\right]| \leq \epsilon$, we have
\begin{align}
\{|f^r_{(p,q)}|\geq \epsilon\} \implies \{|f^r_{(p,q)} - \mathbb{E}\left[f^r_{(p,q)}\right]|\geq 2\epsilon\} 
\end{align}
and, hence,
\begin{align}
\mathbb{P}\left( |f^r_{(p,q)}|\geq \epsilon\right) \leq  \mathbb{P}\left(|f^r_{(p,q)} - \mathbb{E}\left[f^r_{(p,q)}\right]|\geq 2\epsilon\right)\leq e^{-2N\epsilon^2}.
\end{align}
Similarly,
\begin{align}
\mathbb{P}\left( |f^i_{(p,q)}|\geq \epsilon\right) \leq  \mathbb{P}\left(|f^i_{(p,q)} - \mathbb{E}\left[f^i_{(p,q)}\right]|\geq 2\epsilon\right)\leq e^{-2N\epsilon^2}.
\end{align}

Application of the triangular inequality  $|f_{(p,q)}| \leq |f_{(p,q)}^r|+|f_{(p,q)}^i|$ provides
\begin{align}
\mathbb{P}\left( |f_{(p,q)}|\geq \epsilon\right) \leq\mathbb{P}\left( |f^r_{(p,q)}|\geq \frac{\epsilon}{2}\right)+\mathbb{P}\left( |f^i_{(p,q)}|\geq \frac{\epsilon}{2}\right) \leq 2e^{-\frac{N\epsilon^2}{2}}.
\end{align}
Thus,
\begin{align}
\mathbb{P} \left(\mu(\mathbf{A}) \geq \epsilon\right) &= \mathbb{P} \left( \max_{\substack{p \in \llbracket 1, P-1\rrbracket, \\ q \in \llbracket 1, Q-1\rrbracket}} |f_{(p,q)}|\geq \epsilon\right) \\
& \leq \sum_{\substack{p \in \llbracket 1, P-1\rrbracket, \\ q \in \llbracket 1, Q-1\rrbracket}}\mathbb{P} \left( |f_{(p,q)}|\geq \epsilon\right) \\
& \leq (P-1)(Q-1)\mathbb{P} \left( |f_{(p,q)}|\geq \epsilon\right) \\
& \leq 2(P-1)(Q-1)e^{-\frac{N\epsilon^2}{2}}.
\end{align}

Alternatively, we have
\begin{align}
\mathbb{P} \left(\mu(\mathbf{A}) < \epsilon\right) > 1- 2(P-1)(Q-1)e^{-\frac{N\epsilon^2}{2}}.
\end{align}
Next, denote $\epsilon = \frac{1}{2K-1}$. This produces
\begin{align}
\mathbb{P}\left( \mu(\mathbf{A}) <\frac{1}{2K-1}\right) &> 1- \delta \nonumber\\
&\geq 1- 2(P-1)(Q-1)e^{-\frac{N\left(\frac{1}{2K-1}\right)^2}{2}}.
\end{align}
where 
\begin{align}
&\delta\leq 2(P-1)(Q-1)e^{-\frac{N\left(\frac{1}{2K-1}\right)^2}{2}},
\end{align}
or,
\begin{align}
& \quad K \leq \frac{1}{2} \sqrt{\frac{N}{2(\log 2(P-1)(Q-1) -\log \delta)}}+\frac{1}{2}.
\end{align} 

Hence, coherence of the matrix is upper bounded by $\frac{1}{2K-1}$ with probability $1-\delta$. However, the coherence result holds only if the conditions \eqref{eq:mean_condition} are satisfied. Since we assume that $\epsilon = \frac{1}{2K-1}$, next, we derive a relation between $K$ and the probability density of $d_n$s such that \eqref{eq:mean_condition} holds for $\epsilon = \frac{1}{2K-1}$.  
From \eqref{eq:interval}-\eqref{eq:density}, it follows that  
\begin{align}
|\mathbb{E}\left[f_{(p,q)}^r\right]| \leq \frac{1}{\pi |\mathcal{I}|} \quad \text{and} \quad  |\mathbb{E}\left[f_{(p,q)}^i\right]| \leq \frac{1}{\pi |\mathcal{I}|} \label{eq:mean_condition2}.
\end{align}
Therefore, 
\begin{align}
&\frac{1}{\pi |\mathcal{I}|} \leq \frac{1}{2K-1}, \nonumber
\end{align}
or
\begin{align}
& K \leq \frac{\pi |\mathcal{I}|}{2}+\frac{1}{2}. \label{mean_conditon3}
\end{align}
This completes the proof.
\end{IEEEproof}

Theorem~\ref{thm:mp} implies that the number of recoverable targets is $K=\mathcal{O}(\sqrt{\frac{N}{PQ}})$. In case of RSF, a similar result holds but for larger $N$ and grid sizes dependent on $N$ and $M$. These results are analogous to the uniform recovery conditions derived for the spatial CS radar framework proposed in \cite{rossi2014spatial} where, based on the mutual coherence of the sensing matrix, a relationship between the minimum number of antennas and recoverable targets is established.

\section{Numerical Experiments}
\label{sec:numexp}
We now report performance comparisons of RaSSteR with state-of-the-art approaches for various target and interference conditions. We describe these in the following subsections.

\subsection{Comparative State-of-the-Art}
\label{subsec:sota}
We evaluate RaSSteR against the following methods: linear SFW \cite{levanon2004radar}, RSF \cite{huang2018analysis}, conventional chirped or linear frequency modulated (LFM) PDR \cite{peebles1998radar}, and sub-Nyquist radar \cite{barilan2014focusing}. These approaches are quite different in terms of their waveforms. It is not feasible to compare their results by keeping all design parameters identical for these four systems. Therefore, in order to ensure a fair comparison among all systems, we design them such that these systems have the same (coarse or fine, as applicable) range and Doppler resolution as RaSSteR. The fixed native resolution constraint then decides other parameters which differ from system-to-system. The parameters used are listed in Table~\ref{tbl:params}. 
\begin{table}[t]
 \centering
 \caption{System parameters for numerical experiments}
 \label{tbl:params}
\begin{tabular}{|c|c|c|c|c|}
\hline Parameter & RaSSteR & SFW & LFM PDR & Sub-Nyquist Radar \\
\hline Available bandwidth, $B$ & \multicolumn{3}{c|}{$150$ MHz} & $2.5$ MHz\\
\hline$T_{p}$ & $0.4$ $\mu$s & $7.8$ $\mu$s & $0.4$ $\mu$s & $197$ $\mu$s \\
\hline$T$ & \multicolumn{2}{c|}{$62.5$ $\mu$s} & $90$ $\mu$s & $625$ $\mu$s \\
\hline$f_{c}$ & \multicolumn{4}{c|}{$690$ MHz} \\
\hline CPI & \multicolumn{4}{c|}{$0.88$ s} \\
\hline$\Delta f$ & \multicolumn{2}{c|}{$2.5$ MHz} & N/A & N/A \\
\hline N & \multicolumn{2}{c|}{60} & N/A & N/A \\
\hline Transmitted bandwidth & $40\%$ & $100\%$ & $100\%$ & $40\%$ \\
\hline \multicolumn{5}{|c|}{Resolutions}\\
\hline Doppler & \multicolumn{4}{c|}{$0.25$ m/s} \\
\hline (Fine) range & \multicolumn{3}{c|}{$1$ m} & $60$ m\\
\hline Coarse range & \multicolumn{2}{c|}{$60$ m} & \multicolumn{2}{c|}{$\mathrm{N/A}$} \\
\hline
\end{tabular}
\end{table}

Some of the key differences in terms of the design and processing of these systems vis-\`{a}-vis RaSSteR are as follows:
    \begin{description}
        \item [SFW] Keeping the resolutions of SFW and RaSSteR identical, the pulse width of the resulting SFW is longer than the latter. When the targets are moving, the standard SFW processing is unable to estimate Doppler. Therefore, in order to ensure a fair comparison, our SFW processing also employs OMP-based recovery. This approach, in fact, gives SFW an advantage over its traditional implementations. The superior performance of RaSSteR over SFW demonstrated later in this section should be viewed in this context.
        \item [LFM PDR] Our LFM PDR receiver processing is based on conventional pulse compression (matched filter). Since standard LFM PDR does not have the concept of coarse or fine range resolutions, we choose its design parameters to match the fine resolution of RaSSteR. Note that LFM PDR also utilizes all of the available bandwidth.
        \item [Sub-Nyquist radar] Our sub-Nyquist radar implementation is based on \cite{barilan2014focusing} which transmits a fraction of the Nyquist transmit signal bandwidth. We keep the fraction of transmit bandwidth the same across sub-Nyquist radar and RaSSteR. Note that the bandwidth of $2.5$ MHz listed in Table~\ref{tbl:params} for sub-Nyquist radar is for a single pulse. In the presence of interference, we change the transmission of sub-Nyquist radar to its cognitive version \cite{mishra2017performance,mishra2019cognitive} to allow for fair comparison with cognitive RaSSteR.
        \item [RSF] The waveform parameters for RSF are same as RaSSteR except that the former transmits all the frequencies in the bandwidth. Since the advantage of RaSSteR over RSF is evident in the presence of interference, we skip RSF while comparing the delay-Doppler recovery of RaSSteR for individual target scenarios with the other three approaches (SFW, LFM PDR, and sub-Nyquist radar) because a perfect recovery of RaSSteR in the absence of interference certainly implies perfect recovery with RSF (as mentioned earlier in Section~\ref{sec:perfanal}).
    \end{description}

    \subsection{RaSSteR Frequency Selection}
    \label{subsec:freq_sel}
    Let $\mathcal{I}_{N}$ be the set of $N$ unique frequencies randomly chosen from the available carriers set $\mathcal{I}$. Define the effective bandwidth $B_{\mathrm{eff}}$ as
    \begin{align}
    B_{\text {eff }}=\max _{f_{i}, f_{j} \in \mathcal{I}_{N}}\left|f_{i}-f_{j}\right|,
    \end{align}
    and the effective frequency step $\Delta f_{\text {eff }}$ as
    \begin{align}
    \Delta f_{\mathrm{eff}}=\min _{f_{i}, f_{j} \in \mathcal{I}_{N}}\left|f_{i}-f_{j}\right|.
    \end{align}
    In a random draw of the frequencies, it is highly likely that either $f_{0}$ or $f_{M}$ is not included in $\mathcal{I}_{N}$. As a result, in most cases, the effective bandwidth would be smaller than the real bandwidth $B=f_{M}-f_{0}$. When $N \ll M$, the effective bandwidth could be even narrower than half of the real bandwidth. In that case, the measurements become highly correlated and the probability of CS-based perfect recovery decreases. Hence, in practice, the frequencies must be selected in such a way that the effective bandwidth of the selected frequency bins should be as wide as possible. We ensure this in the experiments by including the first and last available frequencies of $\mathcal{I}$.

    The effective frequency step $\Delta f_{\text {eff }}$ is also related to the resolution. If $\mathcal{I}_N$ includes a few consecutive frequencies from $\mathcal{F}$, then the effective $\Delta f$ is as small as possible. So, our random frequency selection strategy comprises the following steps:
    \begin{enumerate}[label=\arabic*)]
        \item Include the lowest and highest frequencies from the available frequency set $\mathcal{I}$ to achieve maximum $B_{\mathrm{eff}}$.
        \item Include a few adjacent frequencies to achieve minimum $\Delta f_{\mathrm{eff}}$.
        \item Draw the rest of the frequencies uniformly at random from $\mathcal{I}$. 
    \end{enumerate}

    \subsection{Performance Metrics}
    \label{subsec:perfmet}
	\begin{figure*}
		\centering
        \includegraphics[width=1.0\textwidth]{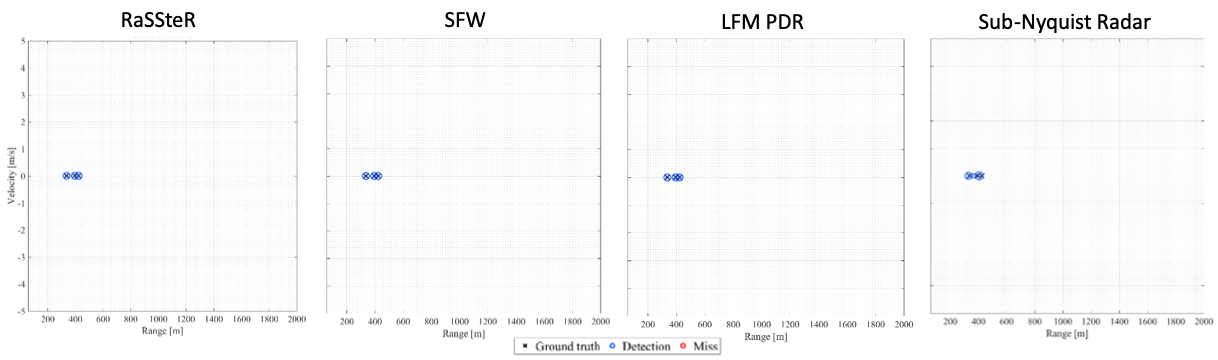}
			\caption{Detection of targets in the delay-Doppler plane for RaSSteR, SFW, LFM PDR, and sub-Nyquist radar for a scenario with widely spaced static targets, K=3, SNR = $-30$ dB, no interference. The black cross indicates the ground truth while blue and red circles imply successful and missed (false alarms included) detections. Successful detection in this case was \textit{sensu stricto}, i.e. the estimated target was at the exact range-Doppler bin of the ground truth.}
			\label{fig:nint_wide_stat}
	\end{figure*}
	\begin{figure*}
		\centering
        \includegraphics[width=1.0\textwidth]{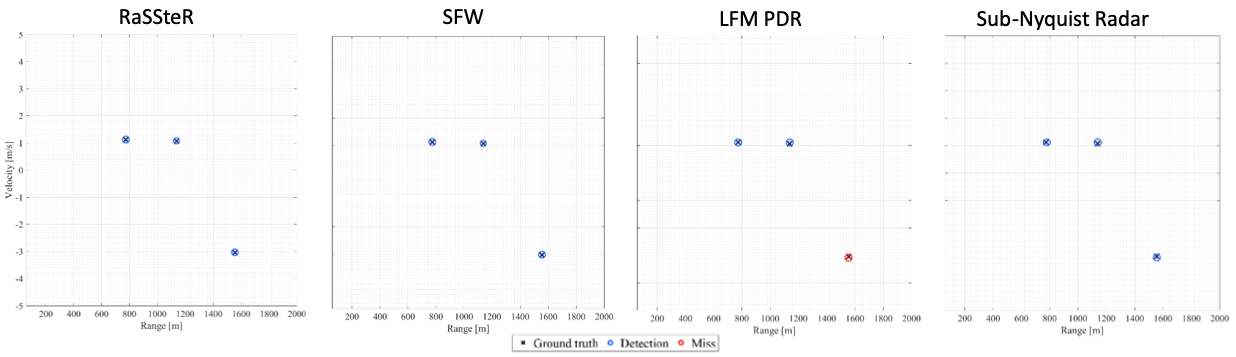}
			\caption{As in Fig.~\ref{fig:nint_wide_stat} but for moving targets.}
			\label{fig:nint_wide_move}
	\end{figure*}
	\begin{figure*}
		\centering
        \includegraphics[width=1.0\textwidth]{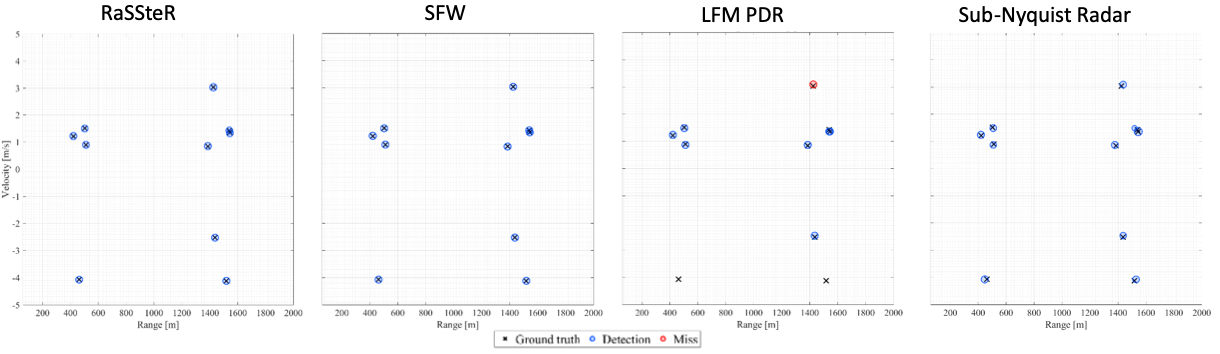}
			\caption{As in Fig.~\ref{fig:nint_wide_stat} but for randomly placed $K=6$ moving targets.}
			\label{fig:nint_rand_move}
	\end{figure*}
	\begin{figure*}
		\centering
        \includegraphics[width=1.0\textwidth]{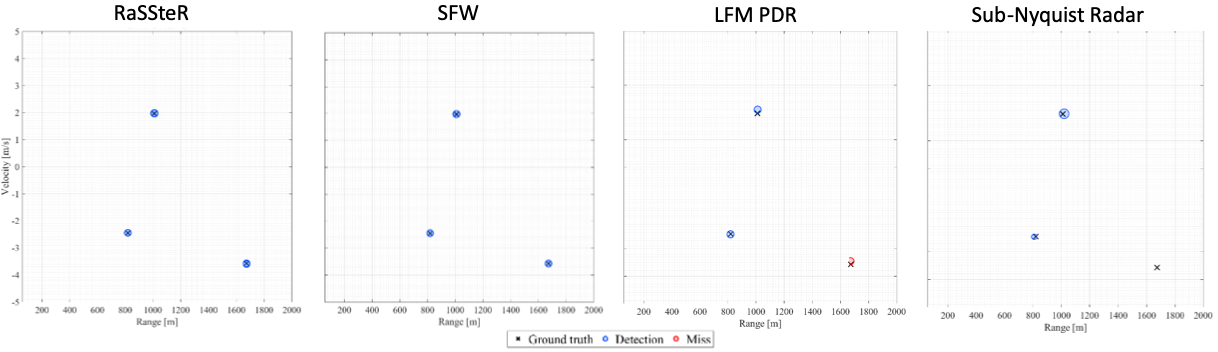}
			\caption{As in Fig.~\ref{fig:nint_wide_stat} but for closely-spaced $K=6$ moving targets ($2$ targets per coarse range bin).}
			\label{fig:nint_clos_move}
	\end{figure*}
	\begin{figure*}
		\centering
        \includegraphics[width=1.0\textwidth]{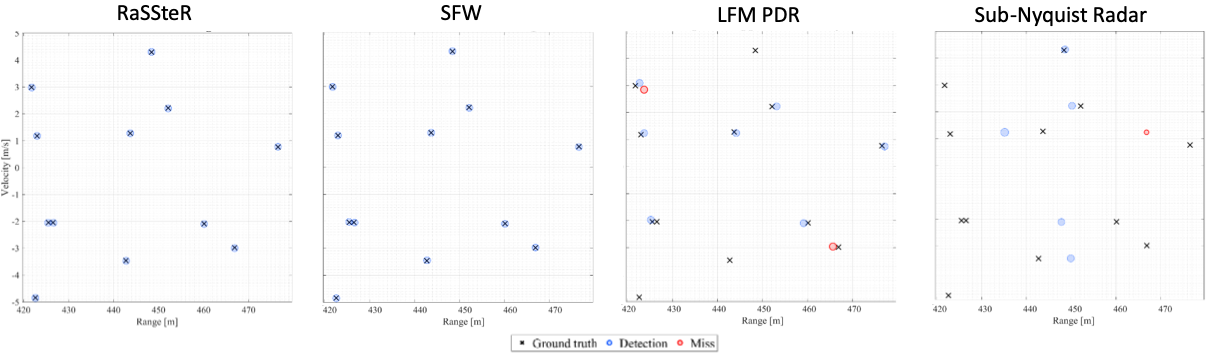}
			\caption{As in Fig.~\ref{fig:nint_wide_stat} but for a dense cluster of $K=12$ moving targets within a single coarse $60$ m range bin of RaSSteR.}
			\label{fig:nint_dens_move}
	\end{figure*}
    We use the following definitions of metrics in our experiments.
    \begin{description}
        \item[SNR] For all experiments, we compute the SNR for the measurement model in \eqref{eq:mes3} as
    \begin{align}
	\text{SNR}= 10 \log_{10}\left(\frac{\|\mathbf{\tilde{y}} \|^2}{N \sigma^2}\right).
	\end{align}
	While comparing different systems, we choose $\sigma$ for the injected noise such that the SNR mentioned above is identical for all the radars.
	    \item[SIR] The interference is modeled as i.i.d circular Gaussian random variable $\mathbf{e} \in \mathbb{C}^N$ with zero mean and variance $\sigma_I^2$. It is added to the measurements $\mathbf{\tilde{y}}$ only for SFW, LFM PDR, and RSF systems because these methods utilize the entire bandwidth. The interference is added to those pulses for which $d_n \in [M_1, M_2]$ resulting in a sparse $\mathbf{e}$. The signal-to-interference-ratio (SIR) is
	    \begin{align}
	    \text{SIR}=10\log_{10} \left(\frac{\|\mathbf{\tilde{y}} \|^2}{ |\Lambda_I| \sigma_I^2}\right),
	    \end{align}
	    where the integer set $\Lambda_I$ is defined as $\Lambda_I=\{n \in \mathbb{Z}: d_n \in [M_1, M_2] \}.$
	\item[Hit rate] In \eqref{eq:mes1}, the target's range and Doppler are defined by  $\{n_k, \nu_k\}_{k=1}^K$. Suppose, for a given SNR and SIR, the estimates of these parameters are given as $\{\hat{n}_k, \hat{\nu}_k\}_{k=1}^K$. Then, we define the (fractional) hit-rate as
	\begin{align}
	\text{HIT RATE} &= \frac{1}{K}  | \{k\in [1, K]: |n_l-\hat{n}_k|\leq T_r \, \nonumber\\
	&\;\;\;\;\;\;\;\;\;\text{and}\,\, |\nu_l-\hat{\nu}_k|\leq T_d, l\in [1, K]\}|,
	\end{align}  
	where $T_r$ and $T_d$ are tolerances in range and Doppler directions, respectively. Unless otherwise stated, $T_r=T_d = 1$ in all experiments. For fixed SNR and SIR, we average hit-rates over $1000$ independent realizations of noise.
	\end{description}
	
	\subsection{Performance in the Absence of Interference}
	\label{subsec:noint}
	We first evaluate RaSSteR processing against other radars in the absence of interference. For static $K=3$ targets that are widely-separated (i.e. all targets in distinct coarse range bins), Fig.~\ref{fig:nint_wide_stat} shows the baseline successful detection performance of all four radars at SNR=-$30$ dB. Then, for the same number of moving targets, each in a different coarse bin, Fig.~\ref{fig:nint_wide_move} shows detections for the same four radar systems. We note that the PC-based processing of LFM PDR has a missed detection. Note that a standard SFW will also perform extremely poorly for moving targets and it is very difficult to estimate Doppler for SFW. Therefore, in order to allow a reasonable comparison of retrieved Doppler, we perform OMP-based recovery for SFW in these experiments. In Fig.~\ref{fig:nint_rand_move}, \textit{ceteris paribus}, we increase the number of targets from $K=3$ to $10$ and randomly placed them across various coarse range bins. The classical LFM PDR continues to underperform. The sub-Nyquist radar performance is on par with the RaSSteR but, as mentioned in Section~\ref{subsec:clust}, the former achieves detection by applying OMP search over a larger grid (across all coarse range bins) than the latter. Next, we consider clustered target detection. For $K=6$ closely spaced targets with $2$ targets per coarse range bin, Fig.~\ref{fig:nint_clos_move} illustrates the high range-resolution advantage offered by SF radars. Here, both classical LFM PDR and sub-Nyquist radar fail to provide $100\%$ detection. Finally, Fig.~\ref{fig:nint_dens_move} shows detection within the $7$-th coarse bin of RaSSteR range profile with a dense cluster of $12$ targets. Note that both RaSSteR and OMP-based SFW provide succesful detection of all targets within this $60$ m HRRP while several missed detections and false alarms are seen in LFM PDR and sub-Nyquist radars.

	\subsection{Cognitive Performance in the Presence of Interference}
	\label{subsec:print}
	\begin{figure*}
		\centering
        \includegraphics[width=1.0\textwidth]{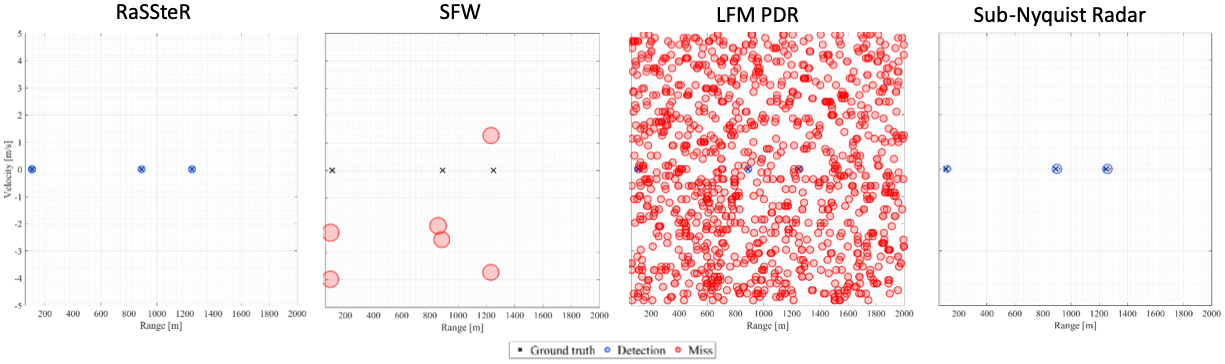}
			\caption{As in Fig.~\ref{fig:nint_wide_stat} but for closely-spaced $K=6$ static targets ($2$ targets per coarse range bin), no AWGN, and SIR=$10$ dB.}
			\label{fig:pint_clos_stat_nless}
	\end{figure*}
	\begin{figure*}
		\centering
        \includegraphics[width=1.0\textwidth]{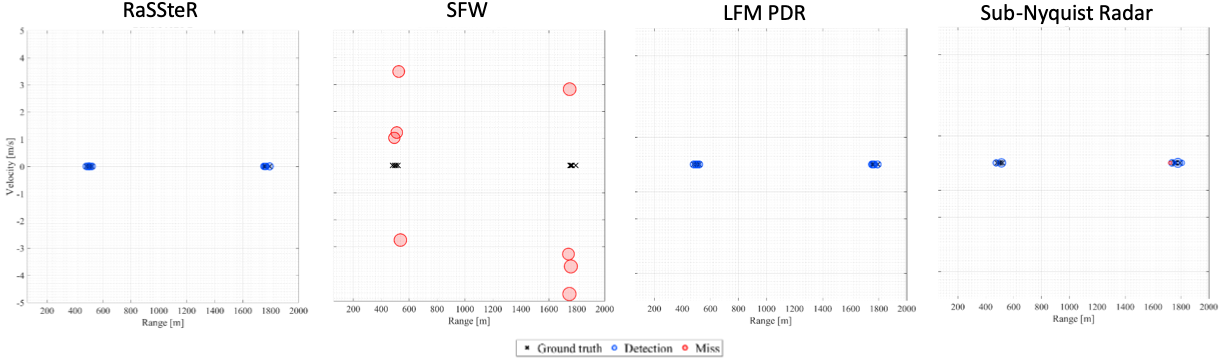}
			\caption{As in Fig.~\ref{fig:pint_clos_stat_nless} but for closely-spaced $K=8$ static targets ($4$ targets per coarse range bin), SNR=-$30$ dB, and SIR=$100$ dB.}
			\label{fig:pint_clos_stat_nfull}
	\end{figure*}
	\begin{figure*}
		\centering
        \includegraphics[width=1.0\textwidth]{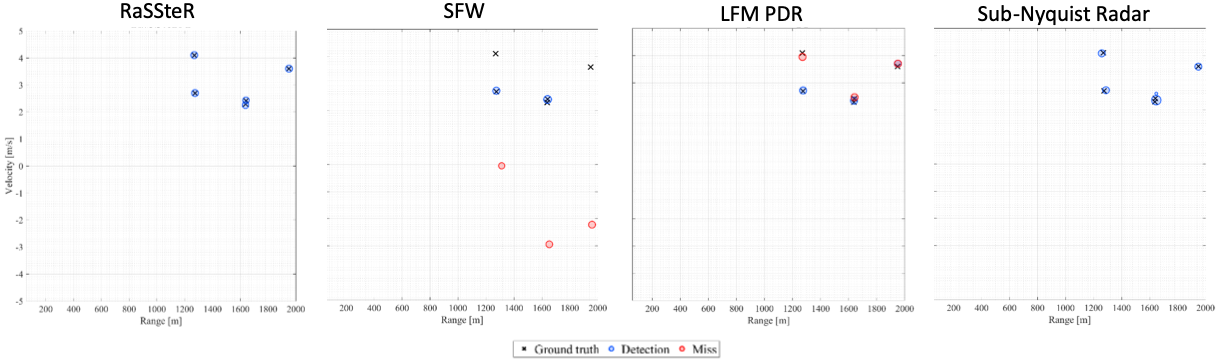}
			\caption{As in Fig.~\ref{fig:pint_clos_stat_nless} but for randomly-spaced $K=5$ moving targets, no AWGN, and SIR=$100$ dB.}
			\label{fig:pint_rand_stat_nless}
	\end{figure*}
	\begin{figure*}
		\centering
        \includegraphics[width=1.0\textwidth]{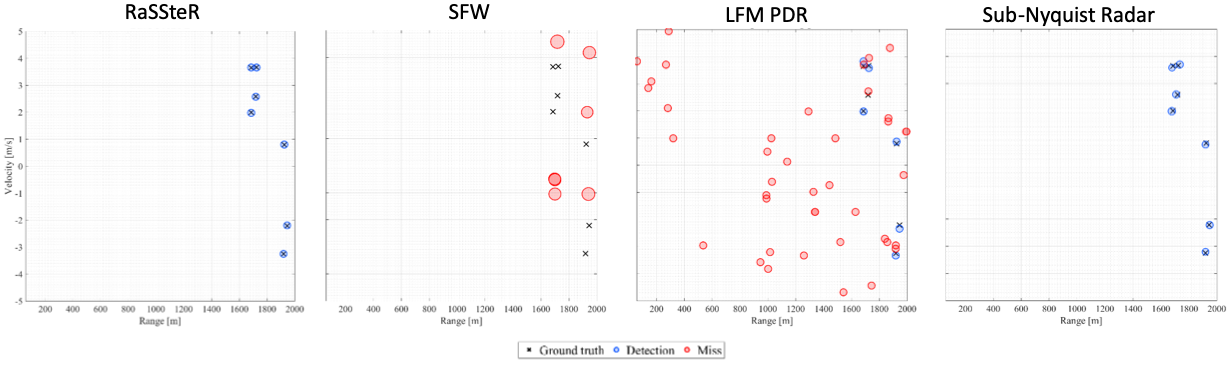}
			\caption{As in Fig.~\ref{fig:pint_rand_stat_nless} but for randomly-spaced $K=7$ moving targets, SNR=-$30$ dB, and SIR=$10$ dB.
			}
			\label{fig:pint_rand_move_nfull}
	\end{figure*}
	\begin{figure*}
		\centering
        \includegraphics[width=1.0\textwidth]{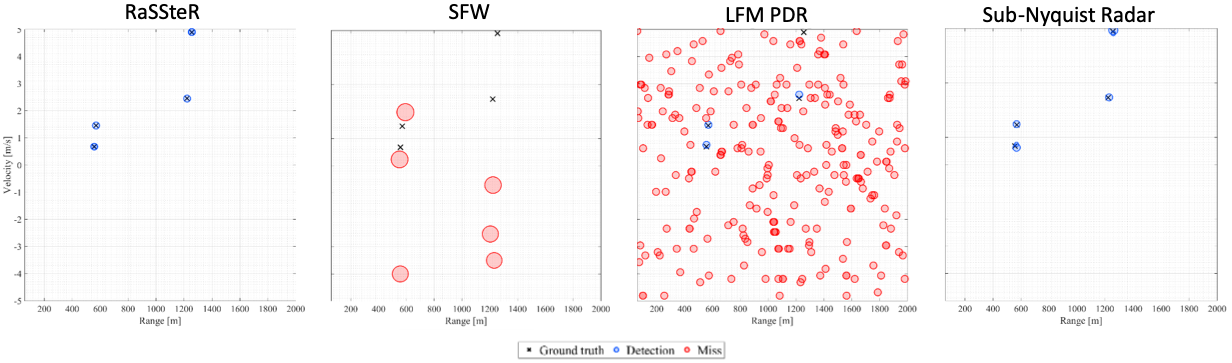}
			\caption{As in Fig.~\ref{fig:nint_wide_stat} but for closely-spaced $K=6$ moving targets (with two coarse bins with $2$ targets each), SNR=-$30$ dB, and SIR=$10$ dB.
			}
			\label{fig:pint_clos_move_nfull}
	\end{figure*}
	\begin{figure}
		\centering
        \includegraphics[width=1.0\columnwidth]{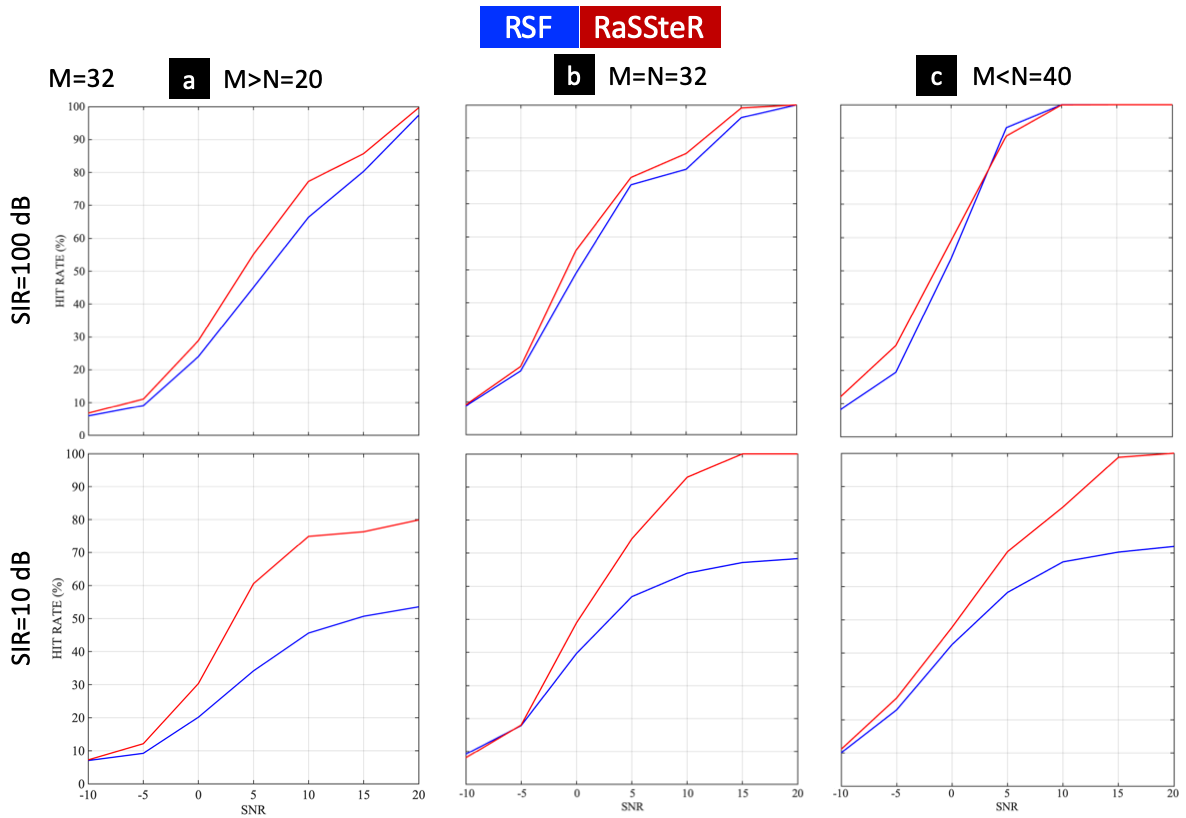}
			\caption{Comparison of RSF and RaSSter for different SNRs and SIR when (a) $N<M$, (b) $N=M$, and (c) $N>M$. For weak interference (SIR $=100$ dB), RaSSteR exhibits a marginally better hit-rate compared than RSF whereas, at SIR $=10$ dB, the improvement is significant.
			\vspace{-10pt}
			}
			\label{fig:hit}
	\end{figure}
	When spectral interference is present, then we employ cognitive transmission for RaSSteR and sub-Nyquist radar so that they avoid the hostile frequencies and transmit with higher per-frequency power than SFW and LFM PDR. For a strong interference scenario of SIR=$10$ dB and no AWGN, Fig.~\ref{fig:pint_clos_stat_nless} shows the baseline performance of the four systems for closely-spaced $K=6$ targets ($2$ per coarse range bin). The cognitive systems (RaSSteR and sub-Nyquist radar) easily outperform the non-cognitive classical SFW and LFM PDR. When the SIR is increased to $100$ dB and AWGN is injected with SNR=-$30$ dB, Fig.~\ref{fig:pint_clos_stat_nfull} shows a scenario with $K=8$ targets that are further clustered (4 per coarse bin). The presence of interference continues to degrade SFW performance. The sub-Nyquist radar shows a few false alarms for this dense static cluster. The interference being very weak, the LFM PDR - whose native resolution of $1$ m is same as the RaSSteR fine range resolution - is able to detect all targets. For $K=5$ moving targets that are randomly located across the delay-Doppler plane, Fig.~\ref{fig:pint_rand_stat_nless} shows the performance in the presence of weak interference without injected AWGN. The cognitive systems again outperform SFW and LFM PDR. For a similar randomly placed target scenario with $K=7$, Fig.~\ref{fig:pint_rand_move_nfull} illustrates the superior performance of cognitive SFW and sub-Nyquist systems when interference is strong (SIR=$10$ dB) and AWGN is injected with SNR=-$30$ dB. When targets are moving, interference is strong, and some targets are closely spaced (a few coarse bins with more than one target), Fig.~\ref{fig:pint_clos_move_nfull} shows that LFM PDR and SFW performance degrades considerably in comparison with RaSSteR.

	\subsection{Statistical Performance}
	\label{subsec:stat_perf}
	Finally, we evaluate the impact of interference on RSF and RaSSteR by computing the hit rates. For this test, we choose $M = 32$, $M_1=14$, $M_2 = 24$, $T_r = T_d =1$, $P = Q = 25$, and $K=4$. Other parameters are same as in Table~\ref{tbl:params}. Figure~\ref{fig:hit} compares the performance of both radars at different SNRs. At low SNRs, we notice marginally better performance of RaSSteR over RSF even when the interference is weak (SIR=$100$ dB). However, when SIR is decreased, the performance of RSF significantly deteriorates, including at high SNRs where the difference between RaSSteR and RSF hit rates is approximately $30$\%. Note that RaSSteR outperforms RSF even when $N<M$. Note that $N \geq M$ implies frequency reuse as mentioned in Section~\ref{subsec:reuse}.

\section{Summary}
\label{sec:summ}
In this paper, we designed a new SF radar waveform drawing on the principles of sparse recovery methods. In order to resolve both range-Doppler coupling and spectrum sharing, we devised a random, sparse SF waveform. Our proposed techniques to estimate targets' Doppler improves upon the traditional SFW and SSFW systems employed for obtaining HRRP of static targets. Theoretical analyses and numerical experiments show that the proposed technique is able to accurately estimates the range and the velocity of the moving target while using less carriers than the RSF system. The cognitive operation of RaSSteR focuses the available burst power in the remaining pulses, thus yielding a better recovery than state-of-the-art for complex scenarios when targets are moving, closely-spaced, and riddled with coexisting spectral interference.

\section*{Acknowledgements}
The authors acknowledge fruitful discussions with Tianyao Huang of Tsinghua University.
\balance
\bibliographystyle{IEEEtran}
\bibliography{main}

\end{document}